\documentclass[prl,twocolumn,showpacs,preprintnumbers,superscriptaddress]{revtex4}
\usepackage{graphicx} 
\usepackage{dcolumn}
\usepackage{bm}
\usepackage{booktabs} 
\usepackage[colorlinks=true,linkcolor=red]{hyperref} 
\usepackage{memhfixc} 

\begin{document}

\title{Real-Time Observation of Reactive Spreading of Gold on Silicon}

\author{Nicola Ferralis}

\email{nferralis@berkeley.edu}

\affiliation{Department of Chemical Engineering, University of California, Berkeley, California 94720, USA}

\author{Farid El Gabaly}
\affiliation{Sandia National Laboratories, Livermore, California 94550, USA}

\author{Andreas K. Schmid}

\affiliation{National Center for Electron Microscopy, Lawrence Berkeley National Laboratory, Berkeley, California 94720, USA}

\author{Roya Maboudian}

\affiliation{Department of Chemical Engineering, University of California, Berkeley, California 94720, USA}

\author{Carlo Carraro}

\affiliation{Department of Chemical Engineering, University of California, Berkeley, California 94720, USA}

\date{December 3, 2009}

\begin{abstract}
The spreading of a bilayer gold film propagating outward from gold clusters, which are pinned to clean Si(111), is imaged in real time by low energy electron microscopy. By monitoring the evolution of the boundary of the gold film at fixed temperature, a linear dependence of the spreading radius on time is found. The measured spreading velocities in the temperature range of $800 < T < 930$~K varied from below 100 pm/s to 50 nm/s. We show that the spreading rate is limited by the reaction to form Au silicide, and the spreading velocity is likely regulated by the reconstruction of the gold silicide that occurs at the interface.  
\end{abstract}


\maketitle

Metal spreading dynamics play a defining role in the growth of semiconductor nanostructures, e.g., by determining pattern fidelity of structures grown by catalyzed vapor-liquid-solid or vapor-solid-solid mechanisms \cite{Carraro_SSR2007}. On a fundamental level, the problem is complicated by its multiscale nature. The equilibrium of a partially wetting microscopic drop on a solid surface often entails the existence of a thin adsorbed film, sometimes of monolayer or even sub-monolayer thickness. Similarly, a spreading drop or thick liquid film is preceded by a thin advancing precursor film \cite{DeGennes_RMP1985}, whose kinetics have been widely debated. Patch-spreading experiments have been carried out at different length- and time scales, ranging from essentially static and macroscopic measurements \cite{Gavrilyuk_PCMS2} to microscopic observations  \cite{Venables_UM1985,Venables_SS1986,Ichinokawa_SS1989,Kirschner_JAP1997,Kirschner_PRB1998,Kirschner_EdTringides1997} under dynamic conditions \cite{Slezak_PRB61}, up to the temperature range where evaporation is a factor \cite{Roos_PRL100}. 

\begin{figure}
\includegraphics{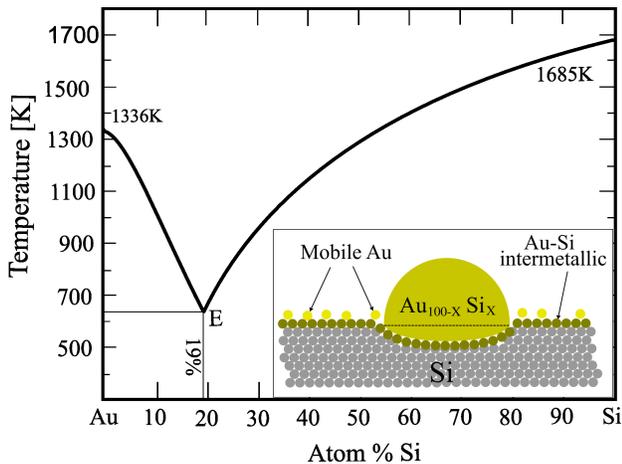}

\caption{\label{Fig1}3D gold/silicon phase diagram. At temperatures above the eutectic T$_E$ =636~K, the composition of the eutectic melt in the cluster is readjusted to follow the eutectic liquidus (bold line). Inset: Schematic diagram of an Au-Si cluster in thermodynamic equilibrium on the Si surface. Under equilibrium condition, the surface is covered by an intermetallic crystalline monolayer of covalently bonded gold silicide with $\sqrt 3\times\sqrt 3$ reconstruction and by additional mobile gold atoms, fed by the Au-Si cluster, with thickness of up to a second monolayer.}

\end{figure}

In this Letter, we employ low-energy electron microscopy (LEEM) to image in real time the evolution of a spreading precursor film
in equilibrium with a cluster reservoir. As in Refs. \cite{Gavrilyuk_PCMS2,Slezak_PRB61} the model system we use is Au/Si(111). In contrast to the previous studies, by focusing on a smaller length scale we address a different limiting regime, as our experimental approach permits us to operate essentially under conditions of constant chemical potential. 
This condition is realized by supplying Au atoms from Au microparticles dispersed on the surface. These particles exist, at elevated temperature, as pinned liquid droplets of Au/Si eutectic melt in bulk (3D) equilibrium with the Si substrate, as shown in Fig.~\ref{Fig1} \cite{Ferralis_JACS2008}. With this experimental method, we are able to determine the non diffusive spreading velocities of the atomically thin precursor film. We show that the linear time dependence in the formation of the interface between the gold silicide and the clean Si surface is a direct consequence of the limited reaction kinetics at the boundary of the spreading precursor film. In atomistic terms, the dynamics of the spreading is regulated by the structural reconstruction of gold silicide that that takes place at the interface.

The experiments were carried out using the spin-polarized LEEM at Lawrence Berkeley National Laboratory \citep{SPLEEM}, which operates under ultrahigh vacuum conditions (base pressure of $5\cdot10^{-11}$ Torr) and on a separate UHV chamber (base pressure of $2\cdot10^{-10}$ Torr), equipped with conventional rear-view low-energy electron diffraction and a cylindrical mirror analyzer for Auger electron spectroscopy \citep{Ferralis_JPCC111}. Gold microspheres (Aldrich, radii between 0.7-1.5~$\mu$m) were dispersed using a N$_2$ carrier gas on hydrogen terminated Si(111) substrates, as described elsewhere \cite{Ferralis_JACS2008}. The areal density of Au clusters was in the range of $10^{-3} \mu m^2$. The Si chips were immediately introduced into ultra-high vacuum. The absence of native oxide was verified by the lack of either SiO$_2$ or O$_2$ peaks in the Auger electron spectra. The heating was performed by electron bombardment of the Si substrate from the back. The temperature during the annealing experiments was measured with a W-Re thermocouple spot-welded to a tantalum plate touching the sample. The thermocouple was calibrated using the Si(111) (1$\times$1) to (7$\times$7) surface phase transition at 1120~K \citep{Bennett_SS104,Slezak_PRB61}, resulting in a precision in the temperature measurement of~$\pm10$~K. Data were recorded with a high resolution charge-coupled device camera, with acquisition rates varying from 1 to 2 frames per second. Data analysis was performed using the image manipulation program ImageJ \citep{ImageJ}.

The 3D gold-silicon phase diagram is characterized by a deep, Au-rich eutectic (composition Au$_{81}$Si$_{19}$ at T$_E$=636~K, Fig.~\ref{Fig1}). This is a well-understood consequence of the frustration in the covalent bonding of silicon brought about by the electron-rich gold. Thus, the solubility of Au in Si is negligible, and the AuSi eutectic melt does not wet the Si surface (the measured contact angle is~$\sim 40^{\circ}$) \cite{Ferralis_JACS2008}. In the dewetting equilibrium, an isolated liquid AuSi drop coexists with a thin adsorbed Au film, uniformly spread over the substrate. This film consists of a crystalline monolayer of gold silicide (i.e., a reconstruction of the Si surface in which Au atoms form three chemical bonds with $\sqrt 3\times\sqrt 3$ structure), upon which a second layer of mobile gold atoms can assemble at high temperature with varying degrees of order \cite{Nogami_PRL65, Nagao_PRB57}.

Upon heating a gold cluster on the H-Si(111), two phenomena occur. First, at T$\sim$T$_E$, the cluster melts, acquiring a composition of Au$_{81}$Si$_{19}$. Previous studies show that the Si is ``dug up" from under the cluster, and the eutectic drop remains pinned in the resulting cavity \cite{Ferralis_JACS2008}. The eutectic drop is in local thermodynamic equilibrium, so that at constant pressure the local chemical potential is specified by the temperature only, along the coexistence curve depicted in bold in Fig.~\ref{Fig1}. Upon raising the temperature above $\sim$783~K, hydrogen desorbs and the surface consists entirely of the clean 7$\times$7 phase with localized eutectic melt drops. Therefore, at any fixed temperature, LEEM allows us to monitor in real time the spreading of the 2D AuSi precursor film, revealed by the contrast between the electron reflectivity of the 7$\times$7 and $\sqrt 3$ reconstructions.

\begin{figure}
\includegraphics{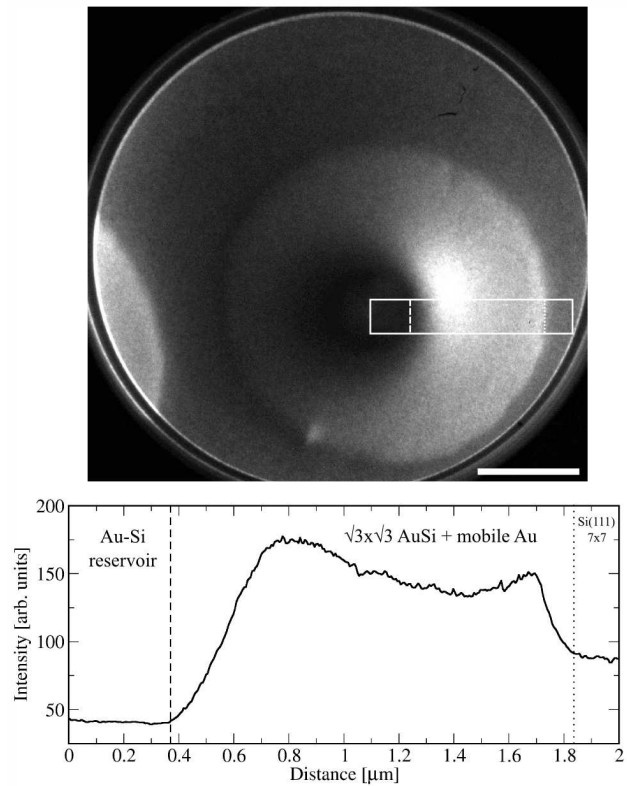}

\caption{\label{Fig2} The increment in the radius of the gold silicide layer is measured directly from the LEEM images. In each LEEM image, the position of the Au-Si eutectic microparticle is within the dark circle~\cite{foot1}, while the surrounding bright disc corresponds to the $\sqrt 3$ reconstructions of the gold silicide layer that forms as a consequence of the reactive spreading of Au over the Si(111) surface. The average radius of this circular spreading annulus is measured from the LEEM images from the average of parallel line profiles (lower plot) acquired from radial rectangular selections, as described in the text. Scale bar:~1~$\mu$m.}

\end{figure}

\begin{figure}
\includegraphics{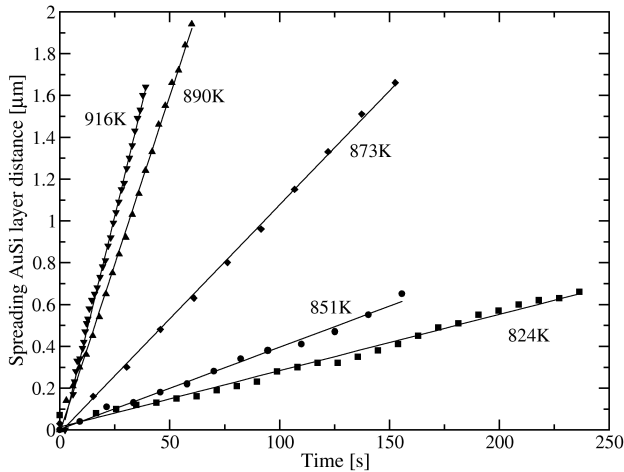}

\caption{\label{Fig3}Gold silicide spreading velocities are measured at different temperatures. The persistence of linear time evolution of the spreading of the silicide layer, incompatible with diffusive spreading, indicates that reaction-limited kinetics endure over a wide temperature range.}
\end{figure}

In a typical experiment, the Si substrate temperature is raised to the desired value (heating rate 5 K/s) and kept fixed while LEEM micrographs are acquired. In each LEEM image (Fig.~\ref{Fig2}), Au-Si droplets appear as dark circles, surrounded by a disc which consists of mobile gold over the gold-silicide layer. For each image, radial boxes are plotted (Fig.~\ref{Fig2}) and radial line profiles within the box are then averaged to improve the signal to noise ratio. The profiles from several radial boxes around the droplet are then averaged. The increment in the radius R(t) of the gold silicide layer is extracted from the resulting profile, as shown in Fig.~\ref{Fig2}, and plots of radius vs time at constant temperature are generated~\cite{foot1}. Several examples are shown in Fig.~\ref{Fig3}. The slopes of the curves show the spreading velocity of the~$\sqrt 3$ surface reconstruction layer spreading outwardly from drops of AuSi eutectic melt at the indicated temperatures. 

In the range of (fixed) temperatures used in these experiments ($800 <T<930$~K), the plots display always a linear behavior, showing that the surface reconstruction spreads at constant velocity (at a given T). We have measured spreading velocities ranging over nearly three orders of magnitude, from below 100 pm/s to 50 nm/s, some examples are plotted in  Fig.~\ref{Fig3}. Linear growth may seem surprising at first glance, as it implies an increasing flux of atoms from the molten drop. However, in our experiments gold droplets can be regarded as practically inexhaustible reservoirs. Although power laws $R(t)\propto t^\alpha$ ($\alpha\le 0.5$) are often encountered in the spreading of precursor films \cite{Kirschner_EdTringides1997,Cazabat_PRL62}, constant spreading velocities have also been predicted in some models \cite{Abraham_PRL65}. Below, we show that the observed time dependence of the spreading is due to the reaction-limited nature of the structural reconstruction of gold silicide that takes place at the interface.

For the general problem of reactive spreading at interfaces, one can imagine two limiting cases. In a diffusion limited case the area of a homogeneous precursor should expand linearly with time, because of the constant flux of atoms supplied by the source. With circular spreading, this corresponds to square-root dependence of the spreading radius on time (as observed for example in Ref.~\cite{Kirschner_EdTringides1997}.) In a reaction limited case, the spreading velocity is limited by the rate at which the structural transition from the pristine substrate surface to the precursor layer reconstruction occurs. This corresponds to a linear dependence of the spreading radius on time. Other limiting cases are conceivable, including the presence of a length scale over which the reaction limited regime becomes diffusion limited, as the distance between the source and the boundary of the precursor layer increases. As discussed below, all our measurements, spanning a wide range of temperatures and spreading velocities, are within the reaction-limited case, although earlier experiments probing Au/Si spreading on a much larger length scale appear to indicate diffusion limited conditions \cite{Gavrilyuk_PCMS2}, suggesting that the crossover length scale is between the regime probed here and that discussed in Ref.~\cite{Gavrilyuk_PCMS2}. 

\begin{figure}
\centerline{\includegraphics[width=0.45\textwidth]{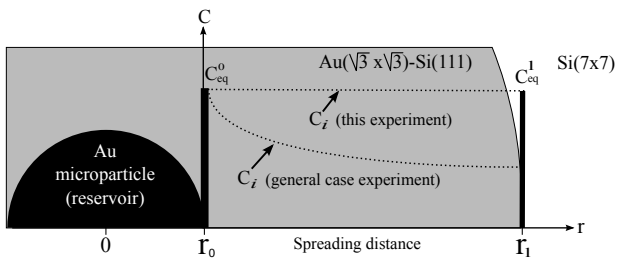}}
\caption{\label{Fig4} Schematics of the evolution of the concentration $\mathrm{C}$ of the Au atoms spreading.}
\end{figure}

In order to understand which mechanisms can be responsible for the non diffusive behavior, let us consider the diffusion equation for steady state conditions that governs the concentration $\mathrm{C_{Au}}$ of Au atoms at the surface (Fig.~\ref{Fig4}),

\begin{equation}
\nabla^2 \mathrm{C}_{\mathrm{Au}} =0.
\label{steady}
\end{equation}
Assuming the Au chemical potential on the terrace in proximity to the Au-Si eutectic microparticle is in equilibrium with bulk Au, the equilibrium concentration of Au atoms close to the microparticle, $C_{eq}^{r_0}$, can be considered constant in time, i.e., $C(r_0)=C_{eq}^{r_0}$, where $C(r_0)$ is the concentration of Au atoms at position $r_0$. In the temperature range used in these experiments, the concentration and mobility of Au adatoms over the Si-($7\times7$) region is very low or zero \cite{Gavrilyuk_PCMS2}; this is also reflected by the lack of notable structural changes in this region prior the formation of the silicide. By solving Eq.~\ref{steady} in cylindrical coordinates, the steady state flux is given by Fick's first law,
\begin{equation}
J_{diff} = -D {\mathrm{d}\mathrm{C_{Au}}\over\mathrm{d} r} = D{{C_{eq}^{r_0}-C(r_1)}\over{\ln({r_1}/{r_0})}}{{1}\over{r}}
\label{velocity_diff}
\end{equation}
where $D$ is the diffusivity of Au atoms over the silicide and $C(r_1)$ the concentration at $r_1$ (see Fig.~\ref{Fig4}). To preserve the steady state condition, Au arriving at the interface must be consumed in the two main mechanisms responsible for the advance of the interface: (i) the formation of the AuSi silicide over the Si surface, and (ii) the rearrangement of Si atoms to reconstruct the surface underneath from Si-($7\times7$) to Si-($\sqrt 3\times\sqrt 3$). We define the reactive flux at the interface as
\begin{equation}
J_{reac} = m [C(r_1)-C_{eq}^{r_1}]  \label{velocity_reac}
\end{equation}
where $C(r_1)-C_{eq}^{r_1}$ corresponds to the deviation from the Au concentration at which both sides of the interface would be in equilibrium \cite{foot2}. $m$ is a constant determined by how fast Au adatoms are incorporated into the interface. Under steady state conditions, both fluxes should be equal, $J_{diff}=J_{reac}$, a condition that leads to a value of $C(r_1)$. Using this value in Eq.~\ref{velocity_diff} we can define an interfacial spreading flux, 
\begin{equation}
J_{interface} = {{m(C_{eq}^{r_0}-C_{eq}^{r_1})}\over{1+{{mr\ln({r_1}/{r_0})}/{D}}}}
\label{diff_reac}
\end{equation}
The factor ${{mr\ln({r_1}/{r_0})}/{D}}$ determines a length scale which specifies when the interface motion crosses over from interface reaction limited ($D\gg {mr\ln({r_1}/{r_0})}$) to diffusion limited ($D\ll {mr\ln({r_1}/{r_0})}$). Note that in the reaction limited case, the concentration gradient across the spreading region is small, $C(r) \sim C_{eq}^{r_0}$ (see Fig.~\ref{Fig4}). If the interface velocity were diffusion-limited, i.e., dominated by the diffusion of Au over the silicide, the measured velocity would be {non-linear} over the measured distance. Furthermore, the variation in the distance the interface moves, $r$, (up to 3~$\mu$m in our experiments, see Fig.~\ref{Fig3}) would result in a non-constant velocity measurement independent of the value of ${D(C_{eq}^{r_0}-C_{eq}^{r_1})}$.

Our observation is that the interface advances at constant velocity for a given temperature, over the mentioned distance. 
This indicates that the interface velocity is reaction limited: the diffusivity of Au over the silicide is so fast that it does not affect the final velocity. The interface flux $J_{interface}$ corresponds to the reaction flux at the interface, $J_{reac}$, which is ultimately controlled by the reaction rate at the interface, (Eq.~\ref{velocity_reac}). This rate is determined by the slowest atomistic processes necessary to move the interface. Since the attachment or detachment of Si from Si steps to reconstruct the surface is the process with higher barrier it is possibly the limiting process.

The length scale at which the crossover to diffusion limited kinetics is expected depends on the value of the diffusivity $D$ of Au on the silicide \cite{Gavrilyuk_PCMS2} and on the Au atom concentration $C$, which cannot be determined from these experiments. The Au atom concentration might be just barely more than the interface reaction can consume (close to crossover), or the availability of gold adatoms could exceed the reaction rate by a huge factor (far from crossover), either case would look identical in these experiments.   
Understanding the mechanism controlling the spreading of Au on Si surfaces is of practical importance in applications where Au clusters are used as catalysts for growth of complex nanostructures, such as epitaxial growth of branched nanotrees by catalyst re-flow \cite{Kawashima_NL8, Doerk_JMC2008}. On a fundamental level, we have shown that the relevance of atomic kinetics in the adlayer structure is dictated by a non-diffusive process. These reaction processes fix the velocity of an otherwise fast diffusing Au for the precursor film mobility of a partially wetting metal-on-semiconductor system. 

\begin{acknowledgments}
We thank N.C. Bartelt for fruitful discussions. This work was supported by the National Science Foundation under Grant EEC-0425914 through the Center of Integrated Nanomechanical Systems, by DARPA N/MEMS Science and Technology Fundamentals Center on Interfacial Engineering for MEMS, and by the National Center for Electron Microscopy, at the Lawrence Berkeley National Laboratory, which is supported by the U.S. Department of Energy under Contract No. DE-AC02-05CH11231.
\end{acknowledgments}


\begin{thebibliography}{40}
\bibitem{foot2} If the Au concentration were to drop lower than $C_{eq}^l$, then the precursor layer would contract, rather than expand.
\end{thebibliography}
\end{document}